\documentclass[10pt, conference, compsocconf]{IEEEtran}
\usepackage{amsfonts}
\let\labelindent\relax
\usepackage{enumitem}
\usepackage{graphicx}
\usepackage{subcaption}
\usepackage{amsmath}
\usepackage{setspace}
\usepackage{algorithm}
\usepackage[]{algpseudocode}
\usepackage{xspace}
\usepackage{fancyhdr}
\ifCLASSINFOpdf
\else
\fi

\newcommand{\StructMatrix}{StructMatrix}

\begin{document}
\title{\LARGE \bf
	\StructMatrix: large-scale visualization of graphs by means of structure detection and dense matrices}
\thispagestyle{headings}
\author{\IEEEauthorblockN{Hugo Gualdron, Robson Cordeiro, Jose F Rodrigues Jr}
\IEEEauthorblockA{
University of Sao Paulo -- ICMC, CS Department\\
Av Trab Sao-carlense, 400, Sao Carlos, SP, Brazil - 13566-590\\
\{gualdron, robson, junio\}@icmc.usp.br}}

\maketitle

\begin{abstract}
Given a large-scale graph with millions of nodes and edges, how to reveal macro patterns of interest, like cliques, bi-partite cores, stars, and chains? Furthermore, how to visualize such patterns altogether getting insights from the graph to support wise decision-making? Although there are many algorithmic and visual techniques to analyze graphs, none of the existing approaches is able to present the structural information of graphs at large-scale. Hence, this paper describes {\StructMatrix}, a methodology aimed at high-scalable visual inspection of graph structures with the goal of revealing macro patterns of interest. {\StructMatrix} combines algorithmic structure detection and adjacency matrix visualization to present cardinality, distribution, and relationship features of the structures found in a given graph. We performed experiments in real, large-scale graphs with up to one million nodes and millions of edges. {\StructMatrix} revealed that graphs of high relevance (e.g., Web, Wikipedia and DBLP) have characterizations that reflect the nature of their corresponding domains; our findings have not been seen in the literature so far. We expect that our technique will bring deeper insights into large graph mining, leveraging their use for decision making.
\end{abstract}


\IEEEpeerreviewmaketitle

\section{Introduction}
\noindent{Large-scale graphs refer to graphs generated by contemporary applications in which users or entities distributed along large geographical areas -- even the entire planet -- create massive amounts of information; a few examples of those are social networks, recommendation networks, road nets, e-commerce, computer networks, client-product logs, and many others. Common to such graphs is the fact that they are made of recurrent simple structures (cliques, bi-partite cores, stars, and chains) that follow macro behaviors of cardinality, distribution, and relationship. Each of these three features depends on the specific domain of the graph; therefore, each of them characterizes the way a given graph is understood.}

While some features of large graphs are detected by algorithms that produce hundreds of tabular data, these features can be better noticed with the aid of visual representations. In fact, some of these features, given their large cardinality, are intelligible, in a timely manner, exclusively with visualization. Considering this approach, we propose {\StructMatrix}, a methodology that combines a highly scalable algorithm for structure detection with a dense matrix visualization. With {\StructMatrix}, we introduce the following contributions: \\
\vspace{-0.3cm}
\begin{enumerate}[noitemsep,nolistsep]
	\item{{\textbf{Methodology:}} we introduce innovative graph processing and visualization techniques to detect macro features of very large graphs;}
	\item{{\textbf{Scalability:}} we show how to visually inspect graphs with magnitudes far bigger than those of previous works;}
	\item{{\textbf{Analysis:}} we analyze relevant graph domains, characterizing them according to the cardinality, distribution, and relationship of their structures.}
\end{enumerate}

The rest of the paper presents related works in Section \ref{sec:related}, the proposed methodology in Section \ref{sec:method}, experimentation in Section \ref{sec:sexperiments}, and conclusions in Section \ref{sec:conclusions}. Table \ref{tbl:symbols} lists the symbols used in our notation.

\section{Related works}
\label{sec:related}

\subsection{Large graph visualization}
\noindent{There are many works about graph visualization, however, the vast majority of them is not suited for large-scale. Techniques that are based on node-link drawings cannot, at all, cope with the needs of just a few thousand edges that would not fit in the display space. Edge bundling \cite{4015425} techniques are also limited since they do not scale to millions of nodes and also because they are able to present only the main connection pathways in the graph, disregarding potentially useful details. Other large-scale techniques are visual in a different sense; they present plots of calculated features of the graph instead of depicting their structural information. This is the case of Apolo \cite{chau2011apolo}, Pegasus \cite{kang2009pegasus}, and OddBall \cite{akoglu2010oddball}. There are also techniques \cite{bertini2004chance} that rely on sampling to gain scalability, but this approach assumes that parts of the graph will be absent; parts that are of potential interest.}

Adjacency matrices in contrast to Node-Link diagrams are the most recommended techniques for fine inspection of graphs in scalable manner \cite{1382886}; this is because they can represent an edge for each pixel in the display. However, even with one edge per pixel, one can visualize roughly a few million edges. Works Matrix Zoom\cite{1382907} and ZAME\cite{4475479} extend the one-edge-per-pixel approach by merging nodes and edges through clustering algorithms, creating an adjacency matrix where each position represents a set of edges on a hierarchical aggregation. The main challenge of using clustering techniques is to find an aggregation algorithm that produces a hierarchy that is meaningful to the user. There are also matrix visualization layouts as MatLink \cite{Henry:2007:MEM:1778331.1778362} and NodeTrix \cite{4376154} combining Node-Link and adjacency representations to increase readability and scalability, but those approaches are not enough to visualize large-scale graphs.

Net-Ray \cite{kang2014net} is another technique working at large scale; it plots the original adjacency matrix of one large graph in the much smaller display space using a simple projection: the original matrix is scaled down by means of straight proportion. This approach causes many edges to be mapped to one same pixel; this is used to generate a heat map that informs the user of how many edges are in a certain position of the dense matrix.

In this work, we extend the approach of adjacency matrices, as proposed by Net-Ray, improving its scalability and also its ability to represent data. In our methodology, we introduce two main improvements: (1) our adjacency matrix is not based on the classic node-to-node representation; we first condense the graph as a collection of smaller structures, defining a structure-to-structure representation that enhances scalability as more information is represented and less compression of the adjacency matrix is necessary; and (2) our projection is not a static image but rather an interactive plotting from which different resolutions can be extracted, including the adjacency matrix with no overlapping -- of course, considering only parts of the matrix that fit in the display. 

\subsection{Structure detection}
\label{sec:structdetec}
\noindent{The principle of {\StructMatrix} is that graphs are made of simple structures that appear recurrently in any graph domain. These structures include cliques, bipartite cores, stars, and chains that we want to identify. Therefore, a given network can be represented in an upper level of abstraction; instead of nodes, we use sets of nodes and edges that correspond to substructures. The motivation here is that analysts cannot grasp intelligible meaning out of huge network structures; meanwhile, a few simple substructures are easily understood and often meaningful. Moreover, analyzing the distribution of substructures, instead of the distribution of single nodes, might reveal macro aspects of a given network.\\}
\vspace{-0.2cm}

\noindent{\textbf{Partitioning (shattering) algorithms}}\\
{\StructMatrix}, hence, depends on a partitioning (shattering) algorithm to work. Many algorithms can solve this problem, like Cross-associations \cite{Chakrabarti04netmine:new}, Eigenspokes \cite{prakash2010eigenspokes}, and METIS \cite{Lasalle2511454}, and VoG \cite{doi:10.1137/1.9781611973440.11}. We verified that VoG overcomes the others in detecting simple recurrent structures considering a limited well-known set.

Vog relies on the technique introduced by graph compression algorithm Slash-Burn by Kang and Faloutsos \cite{DBLP:conf/icdm/KangF11}. The idea of Slash-Burn is that, in contrast to random graphs or lattices, the degree distribution of real-world networks obeys to power laws; in such graphs, a few nodes have a very high degree, while the majority of the nodes have low degree. Kang and Faloutsos also demonstrated that large networks are easily shattered by an ordered ``removal'' of the hub nodes. In fact, after each removal, a small set of disconnected components (satellites) appear, while the majority of the nodes still belong to the giant connected component. That is, the disconnected components were connected to the network only by the hub that was removed and, by progressively removing the hubs, the entire graph is scanned part by part. Interestingly, the small components that appear determine a partitioning of the network that is more coherent than cut-based approaches \cite{LeskovecWWW2008}. The technique works for any power-law graph without domain-specific knowledge or specific ordering of the nodes. 

\begin{figure}
	 \centering
	 \includegraphics[width=200px]{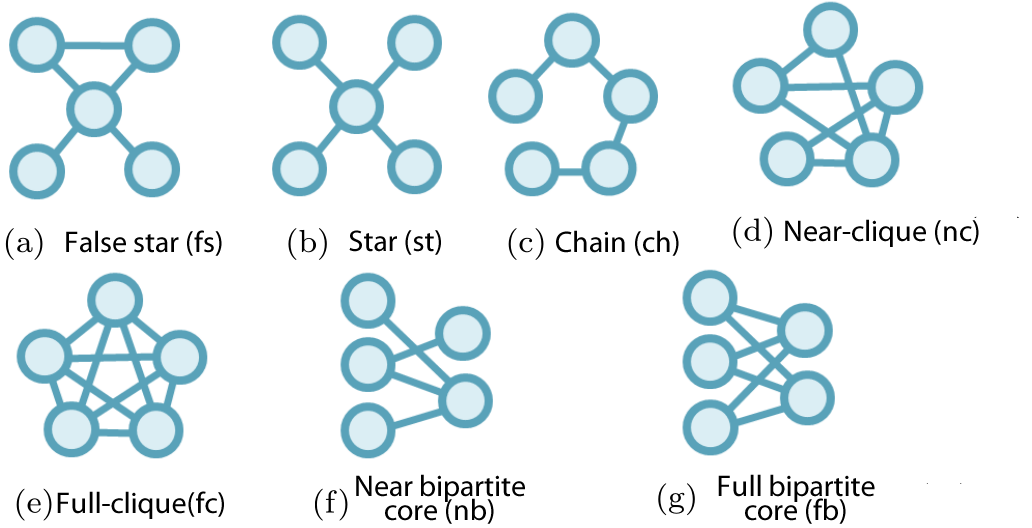}
	\caption{The vocabulary of graph structures considered in our methodology. From (a) to (g), illustrative examples of the patterns that we consider; we process variations on the number of nodes and edges of such patterns.}
	\label{fig:vocabulary}
	\vspace{-0.5cm}
\end{figure}

For the sake of completeness and performance, we designed a new algorithm that, following the Slash-Burn technique, extends algorithm Vog with parallelism, optimizations, and an extended vocabulary of structures, as detailed in Section \ref{sec:algo}. Our results demonstrated better performance while considering a larger set of structures.

\section{Proposed method: {\StructMatrix}}
\label{sec:method}
\noindent{As we mentioned before, \emph{\StructMatrix} draws an adjacency matrix in which each line/column is a structure, not a single node; besides that, it uses a projection-based technique to ``squeeze'' the edges of the graph in the available display space, together with a heat mapping to inform the user of how big are the structures of the graph. In the following, we formally present the technique.}

\begin{table}
	\centering
	\scalebox{0.88}{
	\begin{tabular}{l p{5cm} }
		\hline       
		\textbf{Notation} & \textbf{Description} \\
		\hline  
		$G (V, E)$ & graph with V vertices and E edges\\
		$S, S_{x}$ & structure-set\\	
		$n$, $|S|$ & cardinality of $S$\\
		$M, m_{x,y}$ & StructMatrix\\	
		$fc$, $nc$ & full and near clique resp.\\	
		$fb$, $nb$ & full and near bipartite core resp.\\		
		$st$, $fs$, $ch$ & star, false star and chain resp.\\
		$\psi$ & vocabulary (set) of structures\\
		$D(s_i,s_j)$&Number of edges between structure instances $s_i$ and $s_j$\\
		\hline  
	\end{tabular}
	}
	\caption{Description of the major symbols used in this work.}
	\label{tbl:symbols}
	\vspace{-0.5cm}
\end{table}

\subsection{Overview of the graph condensation approach}
\noindent{For this work, we use a vocabulary of structures that extends those of former works; it considers seven well-known structures -- see Figure \ref{fig:vocabulary} -- found in the graph mining literature: false stars (\emph{fs}), stars (\emph{st}), chains (\emph{ch}), \emph{near} and \emph{full} cliques (\emph{nc}, \emph{fc}), \emph{near} and \emph{full} bi-partite cores (\emph{nb}, \emph{fb}). Shortly, we define the vocabulary of structures as $\psi=\{fs, st, ch, nc, fc, nb, fb\}$.}

\emph{False} stars are structures similar to stars (a central node surrounded by satellites), but whose satellites have edges to other nodes, indicating that the star may be only a substructure of a bigger structure -- see Figure \ref{fig:vocabulary}. A near-clique or $\epsilon$\emph{-near clique} is a structure with $1-\epsilon$ ($0<\epsilon<1$) percent of the edges that a similar full clique would have; the same holds for near bipartite cores. In our case, we are considering $\epsilon = 0.2$ so that a structure is considered \emph{near clique} or \emph{near} bipartite core, if it has at least 80 percent of the edges of the corresponding full structure.

The rationale behind the set of structures $\psi$ is that (a) cliques correspond to strongly connected sets of individuals in which everyone is related to everyone else; cliques indicate communities, closed groups, or mutual-collaboration societies, for instance. (b) Chains correspond to sequences of phenomena/events like those of ``spread the word'', according to which one individual passes his experience/feeling/impression/contact with someone else, and so on, and so forth; chains indicate special paths, viral behavior, or hierarchical processes. (c) Bipartite cores correspond to sets of individuals with specific features, but with complementary interaction; bipartite cores indicate the relationship between professors and students, customers and products, clients and servers, to name a few. And, (d) stars correspond to special individuals highly connected to many others; stars indicate hub behavior, authoritative sites, intersecting paths, and many other patterns.

Considering these motivations, our algorithm condenses the graph in a dense adjacency matrix. To do so, it produces a set with the instances of structures in $\psi$ that were found in the graph; this set of instances contains the same information as that of the original graph but with vertices and edges grouped as structures. Beyond that, the algorithm detects the edges in between the structures, so that it becomes possible to build a condensed adjacency matrix that informs which structure is connected to each other structure.

\subsection{\StructMatrix\ algorithm}
\label{sec:algo}
\noindent{As mentioned earlier, our algorithm is based on a high-degree ordered removal of hub nodes from the graph; the goal is to accomplish an efficient shattering of the graph, as introduced in Section \ref{sec:structdetec}. As we describe in Algorithm \ref{alg:evog}, our process relies on a queue, $\Phi$, which contains the unprocessed connected components (initially the whole graph), and a set $\Gamma$ that contains the discovered structures. In line 4, we explore the fact that the problem is straight parallelizable by triggering threads that will process each connected component in queue $\Phi$. In the process, we proceed with the ordered removal of hubs -- see line 5, which produces a new set of connected components. With each connected component, we proceed by detecting a structure instance in line 7, or else, pushing it for processing in line 10. The detection of structures and the identification of their respective types occur according to Algorithm \ref{alg:vocabulary}, which uses edge arithmetic to characterize each kind of structure.}

\begin{algorithm}
	\caption{\StructMatrix\ algorithm}
	\begin{algorithmic}[1]
		\Require Graph $G = (E, V)$ 
		\Ensure Array $\Gamma$ containing the structures found in $G$
		\State Let be queue $\Phi = \left \{ G \right \}$ and set $\Gamma =  \left \{  \right \}$
		\While{$\Phi$ is not empty}
		\State $H = $Pop($\Phi$) /*Extract the first item from queue $\Phi$*/
		\State SUBFUNCTION Thread($H$) BEGIN /*In parallel*/
		\State $H' = $ ``$H$ without the $1\%$ nodes with highest degree''
		\For{each connected component $cc \in H'$}
		\If {$cc \in \psi$ using Algorithm \ref{alg:vocabulary} }
		\State Add($\Gamma$, cc)
		\Else
		\State Push($\Phi$, cc)
		\EndIf
		\EndFor
		\State END Thread($H$)
		\EndWhile
	\end{algorithmic}
	\label{alg:evog}
\end{algorithm}

\begin{algorithm}
	\caption{Structure classification}
	\begin{algorithmic}[1]
		\Require Subgraph $H = (E, V)$; $n = \left | V \right |$ and $m = \left | E \right |$
		\If {m = $\frac{n(n-1)}{2}$} 
		\Return \emph{fc}
		\ElsIf {$ m > (1-\epsilon) * \frac{n(n-1)}{2}$} %{$ m > \frac{n(n-1)}{2}$}
		\Return \emph{nc}
		\ElsIf {$m < \frac{n^{2}}{4}$ and $H=(E, V_{a} \cup V_{b})$ is bipartite} %($V_{a} \cap V_{b} = \emptyset)$
		\If {$m = \left | V_{a} \right | * \left | V_{b} \right |$}
		\Return \emph{fb}
		\ElsIf {$m > (1-\epsilon) * \left | V_{a} \right | * \left | V_{b} \right |$}
		\Return \emph{nb}
		\ElsIf {$\left | V_{a} \right | =1$ or $\left | V_{b} \right | = 1$}
		\Return \emph{st}
		\ElsIf {$m = n-1$}
		\Return \emph{ch}
		\EndIf
		\EndIf
		\Return undefined structure
	\end{algorithmic}
	\label{alg:vocabulary}
\end{algorithm}

The \StructMatrix\ algorithm, different from former works, maximizes the identification of structures rather than favoring optimum compression; it uses parallelism for improved performance; and considers a larger set of structures. In Section \ref{sec:sexperiments}, we demonstrate these aspects through experimentation. 

\subsection{Adjacency Matrix Layout}
\noindent{A graph $G = \left \langle V, E \right \rangle$ with $V$ vertices and $E$ edges can be expressed as a set of structural instances $S = \{s_{0}, s_{2}, \dots, s_{|S|-1}\}$, where $s_{i}$ is a subgraph of $G$ that is categorized -- see Figure \ref{fig:vocabulary} and Table \ref{tbl:symbols} -- according to the function $type(s): S \rightarrow \psi$. To create the adjacency matrix of structures, first we identify the set $S$ of structures in the graph and categorize each one. Following, we define $n=|S|$ to refer to the cardinality of $S$. 

As depicted in Figure \ref{fig:template}, each type of structure defines a partition in the matrix, both horizontally and vertically, determining subregions in the visualization matrix. In this matrix, a given structure instance corresponds to a horizontal and to a vertical line (w.r.t. the subregions) in which each pixel represents the presence of edges (one or more) between this structure and the others in the matrix. Therefore, the matrix is symmetric and supports the representation of relationships (edges) between all kinds of structure types. Formally, the elements $m_{i,j}$ of a {\StructMatrix} $M_{n\times n}$, $0<i<(n-1)$ and $0<j<(n-1)$ are given by:

\begin{equation}
m_{i,j}=\left\{
\begin{array}{l l}
1,\ if\ D(s_i,s_j) > 0;\\
0\  otherwise.
\end{array}\right.
\end{equation}
where $D:S \times S \rightarrow \mathbb{N}$ is a function that returns the number of edges between two given structure instances. For quick reference, please refer to Table \ref{tbl:symbols}.

\begin{figure}
	\centering
	\includegraphics[height=180px]{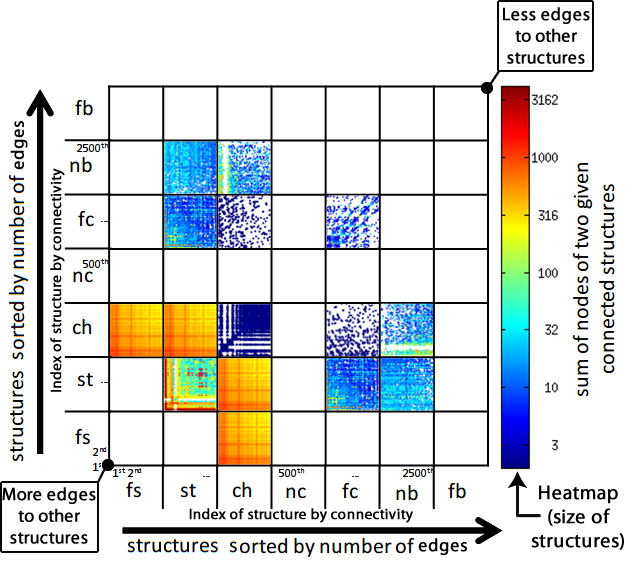}
	\caption{Adjacency Matrix layout.}
	\label{fig:template}
	\vspace{-0.5cm}
\end{figure}

In this work, we focus on large-scale graphs whose corresponding adjacency matrices do not fit in the display. This problem is lessened when we plot the structures-structures matrix, instead of the nodes-nodes matrix. However, due to the magnitude of the graphs, the problem persists. We treat this issue with a density-based visualization for each subregion formed by two types of structures $(\psi_i, \psi_j), \psi_i \in \psi$ and $\psi_j \in \psi$ -- for example, $(fs,fs), (fs,st),...$, and so on. In each subregion, we map each point of the original matrix according to a straight proportion. We map the lower, left boundary point $(x_{min} , y_{min})$ to the center of the lower, left boundary pixel; and the upper, right boundary point $(x_{max}, y_{max})$ to the center of the upper, right boundary pixel. The remaining points are mapped as $(x,y) \rightarrow (\rho_{x},\rho_{y})$ for:

\vspace{-0.3cm}
\begin{equation}
\begin{matrix}
\rho_{x} = R(\psi_{i},\psi_{j})+ \left \lceil (Res_x-1) \frac{x - x_{min}}{x_{max} - x_{min}} + \frac{1}{2}\right \rceil\\  \\
\rho_{y} = R(\psi_{i},\psi_{j})+ \left \lceil (Rex_y-1) \frac{y - y_{min}}{y_{max} - y_{min}} + \frac{1}{2}\right \rceil 
\end{matrix}
\end{equation}
where $R:\psi \times \psi \rightarrow \mathbb{N}$ is a function that returns the offset (left boundary) in pixels of the region $(\psi_{i},\psi_{j})$ and $Res_x, Res_y$ are the target resolutions. The more resolution, the more details are presented, these parameters allow for interactive grasping of details.

Each set of edges connecting two given structures is then mapped to the respective subregion of the visualization where the structures' types cross. Inside each structure subregion we add an extra information by ordering the structure instances according to the number of edges that they have to other structures; that is, by $\sum\limits_{i=0}^{|S|-1} D(s,s_i)$.

Therefore, the structures with the largest number of edges to other structures appear first -- more at the bottom left, less at the top right, of each subregion as explained in Figure \ref{fig:template}.

In the visualization, each horizontal/vertical line (w.r.t. the subregions) corresponds to a few hundred or thousand structure instances; and each pixel corresponds to a few hundred or thousand edges. We deal with that by not plotting the matrix as a static image, but as a dynamic plot that adapts to the available space; hence, it is possible to select specific areas of the matrix and see more details of the edges. It is possible to regain details until reaching parts the original plot, when all the edges are visible.

We plot one last information using color to express the sum of nodes of two given connected structures. We use a color map in which the smaller number of nodes is indicated with bluish colors and the bigger number of nodes is indicated with reddish colors. In addition, we use the same information as used for color encoding to determine the order of plotting: first we plot the edges of the smaller structures (according to the number of nodes), and then the edges of the bigger structures. This procedure assures that the hotter edges will be over the cooler ones, and that the interesting (bigger) structures will be spotted easier. At this point the elements $m_{i,j}$ of a {\StructMatrix} $M_{n\times n}$, $0<i<(n-1)$ and $0<j<(n-1)$ are given by:
\begin{equation}
m_{i,j}=\left\{
\begin{array}{l l}
C(NNodes(s_i) +NNodes(s_j)),\\ if\ D(s_i,s_j) > 0;\\ \\
0\  otherwise.
\end{array}\right.
\end{equation}

where $NNodes:S \rightarrow \mathbb{N}$ is a function that returns the number of nodes of a given structure instance; and $C:\mathbb{N} \rightarrow [0.0,1.0]$ is a function that returns a continuous value between 0.0 (cool blue for smaller structures) and 1.0 (hot red for bigger structures) according to the sum of the number of nodes in the two connected structures. In our visualization, we map the function $C$ to a log scale and then we apply a linear color scale to the data.

\begin{figure}
	\centering
	\begin{subfigure}[b]{150px}
		\includegraphics[width=\linewidth]{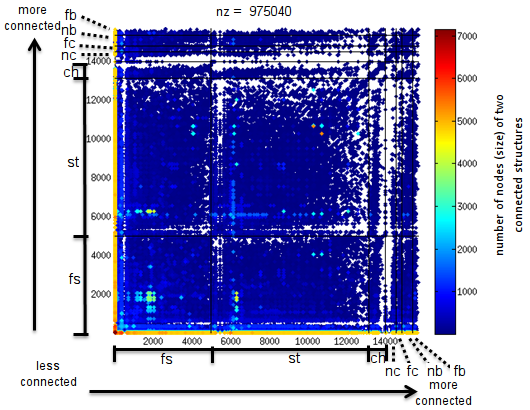}
		\caption{Normal scale.}
		\label{fig:www21}
	\end{subfigure} 
	\begin{subfigure}[b]{150px}
		\includegraphics[width=\linewidth]{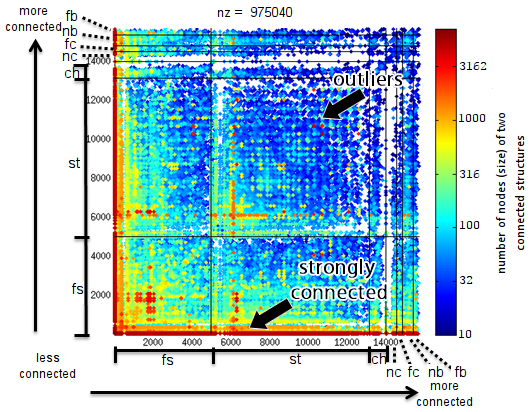}
		\caption{Log scale.}
		\label{fig:www22}
	\end{subfigure} 
	\caption{{\StructMatrix} in the WWW-barabasi graph with colors displaying the sum of the sizes of two connected structures; in the graph, stars refer to websites with links to other websites.}
	\label{fig:www2}
	\centering
	\begin{subfigure}[b]{150px}
		\includegraphics[width=\linewidth]{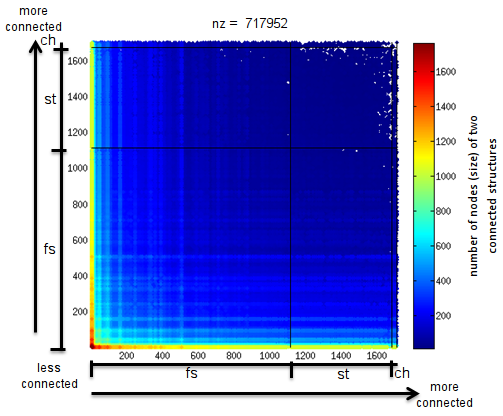}
		\caption{Normal scale.}
		\label{fig:wiki1}
	\end{subfigure} 
	\begin{subfigure}[b]{150px}
		\includegraphics[width=\linewidth]{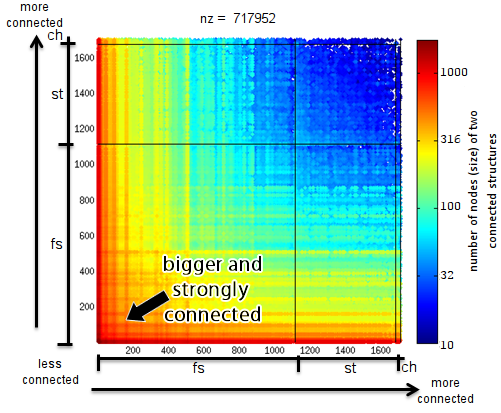}
		\caption{Log scale.}
		\label{fig:wiki2}
	\end{subfigure} 
	\caption{{\StructMatrix} in the Wikipedia-vote graph with values displaying the sum of the sizes of two connected structures; in this graph, stars refer to users who got/gave votes from/to other users.}
	\label{fig:wiki}
		\vspace{-0.5cm}
\end{figure}

\section{Experiments}
\label{sec:sexperiments}
\noindent{Table \ref{tbl:graphs} describes the graphs we use in the experiments.}
\begin{table}[H]
	\centering
		\begin{tabular}{l r r r }
			\hline       
			\textbf{Name} & \textbf{Nodes} & \textbf{Edges} &\textbf{Description} \\
			\hline  
			DBLP 			& 1,366,099 	& 5,716,654 	& Collaboration network \\
			Roads of PA 	& 1,088,092 	& 1,541,898 	& Road net of Pennsylvania\\
			Roads of CA 	& 1,965,206 	& 2,766,607 	& Road net of California\\
			Roads of TX 	& 1,379,917 	& 1,921,660 	& Road net of Texas\\
			WWW-barabasi 	& 325,729 	& 1,090,108 	& WWW in nd.edu\\
			Epinions 		& 75,879 		& 405,740		& Who-trusts-whom network\\
			cit-HepPh 		& 34,546 		& 420,877 	& Co-citation network\\
			Wiki-vote 	& 7,115 		& 100,762 	& Wikipedia votes\\
			\hline  
		\end{tabular}

	\caption{Description of the graphs used in our experiments.}
	\label{tbl:graphs}
	\vspace{-0.4cm}
\end{table}

\subsection{Graph condensations}
\noindent{Table \ref{tbl:summaries} shows the condensation results of the structure detection algorithm over each dataset, already considering the extended vocabulary and structures with minimum size of 5 nodes -- less than 5 nodes could prevent to tell apart the structure types. The columns of the table indicate the percentage of each structure identified by the algorithm. For all the datasets, the false star was the most common structure; the second most common structure was the star, and then the chain, especially observed in the road networks. The improvement of the visual scalability of \emph{\StructMatrix}, compared to former work Net-Ray, is as big as the amount of information that is ``saved'' when a graph is modeled as a structure-to-structure adjacency matrix, instead of a node-to-node matrix.}

\begin{table*}
	\centering
	\scalebox{0.88}{
		\begin{tabular}{ l r r r r r r r r r r}
			\hline       
			\textbf{Graph}  & \textbf{fs} & \textbf{st} & \textbf{ch} & \textbf{nc} & \textbf{fc} & \textbf{nb} & \textbf{fb}\\
			\hline  
			DBLP 			& 122,983	 (76\%)	& 7,585(5\%)		& 3,096(2\%)	& 2,656(2\%)	& 24,551(15\%)	& 14($<$1\%) & -\\
			WWW-barabasi 	& 4,957(32\%)		& 8,146(52\%)	& 851(5\%)	& 541(3\%)	& 283(2\%)	& 556(4\%)	& 318(2\%)\\
			cit-HepPh 		& 11,449(79\%)	& 1,948(13\%)	& 840(6\%)	& 120(1\%)	& 44	(4$<$1\%)	& 35($<$1\%)		& 43($<$1\%)\\
			Wikipedia-vote 	& 1,112(65\%)		& 564(33\%)	& 29	(2\%)	& - 	& - 	& 1($<$1\%) 		&   - 	& 	& \\
			Epinions	 	& 4,518(52\%)		&2,725(31\%) 	& 1,247(14\%)	& 28	(\%)&21(\%) & 150(2\%)	&   3($<$1\%) 	\\
			Roadnet PA 	& 11,825(23\%)	& 22,934(45\%)	& 13,748(27\%)	& -	& -	& 2,668(5\%)	&   -	\\
			Roadnet CA 	& 24,193(27\%)	& 34,781(39\%)	& 26,236(29\%)	& -	& -	& 3,763(4\%)	&   -	\\
			Roadnet TX 	& 15,595(25\%)	& 27,094(43\%)	& 17,457(28\%)	& -	& -	& 2,468(4\%) 	&   - 	\\		
			\hline  
		\end{tabular}
	}
	\caption{Structures found in the datasets considering a minimum size of 5 nodes.}
	\label{tbl:summaries}
	\vspace{-0.5cm}
\end{table*}

\vspace{-0.05cm}
\subsection{Scalability}
\begin{figure}
	\centering
	\includegraphics[width=240px]{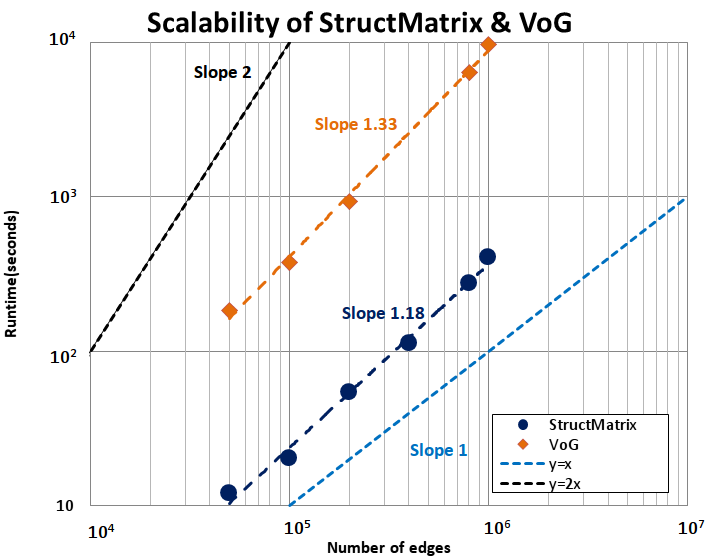}
	\caption{Scalability of the \emph{\StructMatrix} and VoG techniques; although VoG is near-linear to the graph edges, StructMatrix overcomes VoG for all the graph sizes.}
	\label{fig:dblpscalability}
	\vspace{-0.5cm}
\end{figure}	
	
\noindent{In order to test the processing scalability of \emph{\StructMatrix}, we used a breadth-first search over the DBLP dataset to induce subgraphs of different sizes -- we created graphs ranging from 50K edges up to 1.000K edges. For the scalability experiment, we used a contemporary commercial desktop (Intel i7 with 8 GB RAM). We compared the performance between \emph{VoG} and \emph{\StructMatrix} to detect simple recurrent structures from a limited well-known set. Figure \ref{fig:dblpscalability} shows that StructMatrix and VoG are near-linear on the number of edges of the input graph, however StructMatrix overcomes VoG for all the graph sizes.}

\vspace{-0.02cm}
\subsection{WWW and Wikipedia}
\noindent{In Figures \ref{fig:www2} and \ref{fig:wiki}, one can see the results of {\StructMatrix} for graphs WWW-barabasi (325,729 nodes and 1,090,108 edges) and Wikipedia-vote (7,115 nodes and 100,762 edges) condensed as described in Table \ref{tbl:summaries}. For graph WWW-barabasi, Figure \ref{fig:www21} shows the {\StructMatrix} with linear color encoding, and Figure \ref{fig:www22} shows the {\StructMatrix} with logarithmic color encoding. For the Wikipedia-vote graph, the same visualizations are presented in Figures \ref{fig:wiki1} and \ref{fig:wiki2}. We observe the following factors in the visualizations:}

\begin{itemize}[noitemsep,nolistsep]
	\item{the share of structures: WWW-barabasi presents a clear majority of stars, followed by false stars, and chains, while the Wikipedia-vote presents a majority of false stars, followed by stars, and chains; in both cases, stars strongly characterize each domain, as expected in websites and in elections;}
	\item{the presence of outliers in WWW-barabasi, spotted in red; and the presence of structures globally and strongly connected in Wikipedia-vote, depicted as reddish lines across the visualization;}
	\item{the notion that the bigger the structures, the more connected they are -- reddish (the bigger) structures concentrate on the left (the more connected), especially perceived in Wikipedia-vote;}
	\item{the effect of the logarithmic color scale; its use results in a clearer discrimination of the magnitudes of the color-mapped values, what helps to perceive the distribution of the values; more skewed in WWW and more uniform in Wikipedia.}
\end{itemize}

The stars and false stars of the WWW graph in Figure \ref{fig:www22} refer to sites with multiple pages and many out-links -- bigger sites are reddish, more connected sites to the left. The visualization is able to indicate the big stars (sites) that are well-connected to other sites (reddish lines), and also the big sites that demand more connectivity -- reddish isolated pixels. The chains indicate site-to-site paths of possibly related semantics, an occurrence not so rare for the WWW domain. There is also a set of reasonably small, interconnected sites that connect only with each other and not with the others -- these sites determine blank lines in the visualization and their sizes are noticeable in dark blue at the bottom-left corner of the star-to-star subregion. Such sites should be considered as outliers because, although strongly connected, they limit their connectivity to a specific set of sites.

While the Wikipedia graph is mainly composed of stars, just like the WWW graph, the Wikipedia graph is quite different. Its structures are more interconnected defining a highly populated matrix. That means that users (contributors) who got many votes to be elected as administrators in Wikipedia, also voted in many other users. The sizes of the structures, indicated by color, reveal the most voted users, positioned at the bottom-left corner -- the color pretty much corresponds to the results of the elections: of the 2,794 users, only 1,235 users had enough votes to be elected administrators (nearly $50\%$ of the reddish area of the matrix). There are also a few chains, most of them connected to stars (users), especially the most voted ones -- it becomes evident that the most voted users also voted on the most voted users. This is possibly because, in Wikipedia, the most active contributors are aware of each other.

\subsection{Road networks}
\noindent{On the road networks, if we consider the stars segment (``st''), each structure corresponds to a city (the intersecting center of the star); therefore, the horizontal/vertical lines of pixels correspond to the more important cities that act as hubs in the road system. Its \emph{\StructMatrix} visualization -- Figure \ref{fig:roadvisualization} -- showed an interesting pattern for all the three road datasets: in the figure, one can see that the relationships between the road structures is more probable in structures with similar connectivity. This fact is observable in the curves (diagonal lines of pixels) that occur in the visualization -- remember that the structures are first ordered by type into segments, and then by their connectivity (more connected first) in each segment.}

Another interesting fact is the presence of some structures heavily connected to nearly all the other structures; these structures define horizontal lines of pixels in the visualization and, due to symmetry, they also define vertical lines of pixels. The same patterns were observed for roadnets from California, Texas, and Pennsylvania. According to the visualizations, roads are characterized by three patterns:

\begin{enumerate}[noitemsep,nolistsep]
	\item{cities that connect to most of the other cities acting as interconnecting centers in the road structure; these cities are of different importance and occur in small number -- around 6 for each state that we studied;}
	\item{there is a hierarchical structure dictated by the connectivity (importance) of the cities; in this hierarchy, the connections tend to occur between cities with similar connectivity; one consequence of this fact is that going from one city to some other city may require one to first ``ascend'' to a more connected city; actually, for this domain, the lines of pixels in the visualization correspond to paths between cities, passing through other cities -- the bigger the inclination of the line, the shorter the path (the diagonal is the longest path);}
	\item{road connections that are out of the hierarchical pattern -- the ones that do not pertain to any line of pixels; such connections refer to special roads that, possibly, were built on specific demands, possibly not obeying to the general guidelines for road construction.}
\end{enumerate}

\begin{figure}
	\centering
	\begin{subfigure}[b]{160px}
		\centering
		\includegraphics[height=130px]{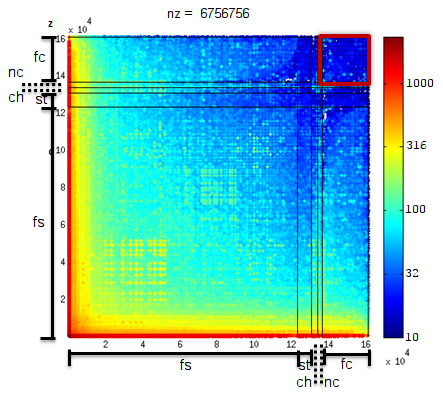}
		\caption{All types of structures.}
		\label{fig:dblp1}
	\end{subfigure} 
	~
	\begin{subfigure}[b]{160px}
		\centering
		\includegraphics[height=130px]{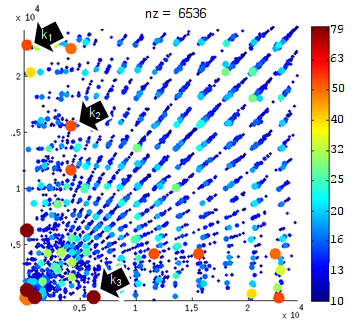}
		\caption{Only the \emph{fc-fc} sub region with details.}
		\label{fig:dblp3}
	\end{subfigure} 
	\caption{DBLP Zooming on the \emph{full} clique section.}
	\label{fig:dblp}
	\vspace{-0.5cm}
\end{figure}

\begin{figure*}
	\centering
	\includegraphics[height=135px]{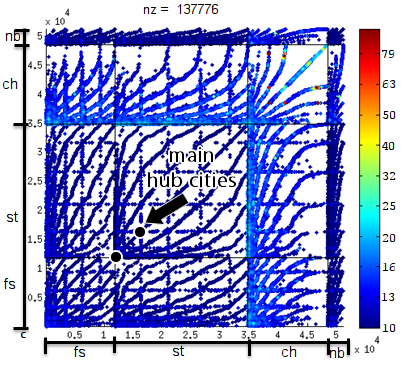}
	\includegraphics[height=135px]{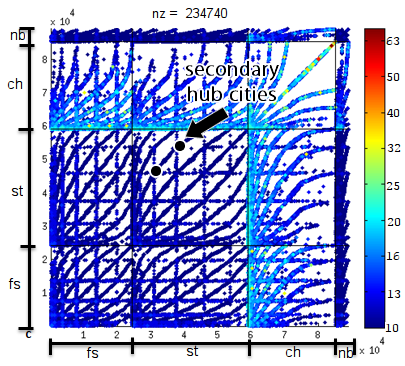}
	\includegraphics[height=135px]{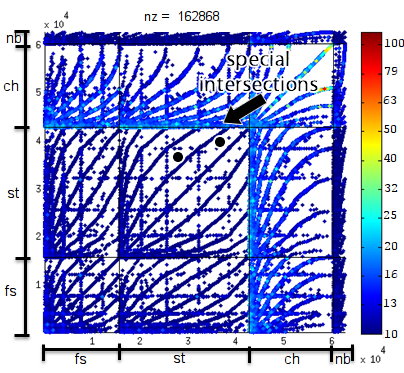}
	\caption{{\StructMatrix} with colors in log scale indicating the size of the structures interconnected in the road networks of Pennsylvania (PA), California (CA) and Texas(TX). Again, stars appear as the major structure type; in this case they correspond to cities or to major intersections.}
	\label{fig:roadvisualization}
	\vspace{-0.5cm}
\end{figure*}

From these visualizations and patterns, we notice that the {\StructMatrix} visualization is a quick way (seconds) to represent the structure of graphs on the order of million-nodes (intersections) and million-edges (roads). For the specific domain of roads, the visualization spots the more important cities, the hierarchy structure, outlier roads that should be inspected closer, and even, the adequacy of the roads' inter connectivity. This last issue, for example, may indicate where there should be more roads so as to reduce the pathway between cities.

\subsection{DBLP}
\noindent{In the \emph{\StructMatrix} of the DBLP co-authoring graph -- see Figure \ref{fig:dblp1} -- it is possible to see a huge number of false stars. This fact reflects the nature of DBLP, in which works are done by advisors who orient multiple students along time; these students in turn connect to other students defining new stars and so on. A minority of authors, as seen in the matrix, concerns authors whose students do not interact with other students defining stars properly said. The presence of full cliques (\emph{fc}) is of great interest; sets of authors that have co-authorship with every other author. Full cliques are expected in the specific domain of DBLP because every paper defines a full clique among its authors -- this is not true for all clique structures, but for most of them. }

In Figure \ref{fig:dblp3}, we can see the full clique-to-full clique region in more details and with some highlights indicated by arrows. The Figure highlights some notorious cliques: $k_{1}$ refers to the publication with title ``\emph{A 130.7mm 2-layer 32Gb ReRAM memory device in 24nm technology}" with 47 authors; $k_{2}$ refers to paper ``\emph{PRE-EARTHQUAKES, an FP7 project for integrating observations and knowledge on earthquake precursors: Preliminary results and strategy}" with 45 authors; and $k_{3}$ refers to paper ``\emph{The Biomolecular Interaction Network Database and related tools 2005 update}" with 75 authors. These specific structures were noticed due to their colors, which indicate large sizes. Structures $k_{1}$ and $k_{3}$, although large, are mostly isolated since they do not connect to other structures; $k_{2}$, on the other hand, defines a line of pixels (vertical and horizontal) of similarly colored dots, indicating that it has connections to other cliques.

\section{Conclusions}
\label{sec:conclusions}
\noindent{We focused on the problem of visualizing graphs so big that their adjacency matrices demand much more pixels than what is available in regular displays. We advocate that these graphs deserve macro analysis; that is, analysis that reveal the behavior of thousands of nodes altogether, and not of specific nodes, as that would not make sense for such magnitudes. In this sense, we provide a visualization methodology that benefits from a graph analytical technique. Our contributions are:}

\begin{itemize}[noitemsep,nolistsep]
	\item{{\textbf{Visualization technique:}} we introduce a processing and visualization methodology that puts together algorithmic techniques and design in order to reach large-scale visualizations;}
	\item{{\textbf{Analytical scalability:}} our technique extends the most scalable technique found in the literature; plus, it is engineered to plot millions of edges in a matter of seconds;}
	\item{{\textbf{Practical analysis:}} we show that large-scale graphs have well-defined behaviors concerning the distribution of structures, their size, and how they are related one to each other; finally, using a standard laptop, our techniques allowed us to experiment in real, large-scale graphs coming from domains of high impact, i.e., WWW, Wikipedia, Roadnet, and DBLP.}
\end{itemize}

Our approach can provide interesting insights on real-life graphs of several domains answering to the demand that has emerged in the last years. By converting the graph's properties into a visual plot, one can quickly see details that algorithmic approaches either would not detect, or that would be hidden in thousand-lines tabular data.

{\footnotesize
	\section*{Acknowledgments}
	\noindent{{\bf We thank Prof. Christos Faloutsos and Dr. Danai Koutra, from Carnegie Mellon University, for their valuable collaboration.} Furthermore, this work received support from Conselho Nacional de Desenvolvimento Cientifico e Tecnologico (CNPq-444985/2014-0), Fundacao de Amparo a Pesquisa do Estado de Sao Paulo (FAPESP-2011/13724-1, 2013/03906-0, 2014/07879-0,  2014/21483-2), and Coordenacao de Aperfeicoamento de Pessoal de Nivel Superior (Capes).}
}
\vspace{-0.2cm}
{	\small
	\newcommand{\BIBdecl}{\setlength{\itemsep}{0.25 em}}
	\bibliographystyle{IEEEtran}

\begin{thebibliography}{10}
\providecommand{\url}[1]{#1}
\csname url@samestyle\endcsname
\providecommand{\newblock}{\relax}
\providecommand{\bibinfo}[2]{#2}
\providecommand{\BIBentrySTDinterwordspacing}{\spaceskip=0pt\relax}
\providecommand{\BIBentryALTinterwordstretchfactor}{4}
\providecommand{\BIBentryALTinterwordspacing}{\spaceskip=\fontdimen2\font plus
\BIBentryALTinterwordstretchfactor\fontdimen3\font minus
  \fontdimen4\font\relax}
\providecommand{\BIBforeignlanguage}[2]{{%
\expandafter\ifx\csname l@#1\endcsname\relax
\typeout{** WARNING: IEEEtran.bst: No hyphenation pattern has been}%
\typeout{** loaded for the language `#1'. Using the pattern for}%
\typeout{** the default language instead.}%
\else
\language=\csname l@#1\endcsname
\fi
#2}}
\providecommand{\BIBdecl}{\relax}
\BIBdecl

\bibitem{4015425}
D.~Holten, ``Hierarchical edge bundles: Visualization of adjacency relations in
  hierarchical data,'' \emph{IEEE TVCG}, vol.~12, no.~5, pp. 741--748, 2006.

\bibitem{chau2011apolo}
D.~H. Chau, A.~Kittur, J.~I. Hong, and C.~Faloutsos, ``Apolo: making sense of
  large network data by combining rich user interaction and machine learning,''
  in \emph{SIGCHI Conf on Human Factors in Computing Systems}, 2011, pp.
  167--176.

\bibitem{kang2009pegasus}
U.~Kang, C.~E. Tsourakakis, and C.~Faloutsos, ``Pegasus: A peta-scale graph
  mining system implementation and observations,'' in \emph{Data Mining, 2009.
  ICDM'09. Ninth IEEE Int Conf on}.\hskip 1em plus 0.5em minus 0.4em\relax
  IEEE, 2009, pp. 229--238.

\bibitem{akoglu2010oddball}
L.~Akoglu, M.~McGlohon, and C.~Faloutsos, ``Oddball: Spotting anomalies in
  weighted graphs,'' in \emph{Adv. in Knowledge Discovery and Data
  Mining}.\hskip 1em plus 0.5em minus 0.4em\relax Springer, 2010, pp. 410--421.

\bibitem{bertini2004chance}
E.~Bertini and G.~Santucci, ``By chance is not enough: preserving relative
  density through nonuniform sampling,'' in \emph{Information
  Visualisation}.\hskip 1em plus 0.5em minus 0.4em\relax IEEE, 2004, pp.
  622--629.

\bibitem{1382886}
M.~Ghoniem, J.~Fekete, and P.~Castagliola, ``A comparison of the readability of
  graphs using node-link and matrix-based representations,'' in \emph{IEEE
  InfoVis}, 2004, pp. 17--24.

\bibitem{1382907}
J.~Abello and F.~van Ham, ``Matrix zoom: A visual interface to semi-external
  graphs,'' in \emph{IEEE InfoVis}, 2004, pp. 183--190.

\bibitem{4475479}
N.~Elmqvist, T.-N. Do, H.~Goodell, N.~Henry, and J.~Fekete, ``Zame: Interactive
  large-scale graph visualization,'' in \emph{PacificVIS}, 2008, pp. 215--222.

\bibitem{Henry:2007:MEM:1778331.1778362}
N.~Henry and J.-D. Fekete, ``Matlink: Enhanced matrix visualization for
  analyzing social networks,'' in \emph{Int Conf on Human-computer
  Interaction}.\hskip 1em plus 0.5em minus 0.4em\relax Springer-Verlag, 2007,
  pp. 288--302.

\bibitem{4376154}
N.~Henry, J.~Fekete, and M.~J. McGuffin, ``Nodetrix: a hybrid visualization of
  social networks,'' \emph{IEEE TVCG}, vol.~13, no.~6, pp. 1302--1309, 2007.

\bibitem{kang2014net}
U.~Kang, J.-Y. Lee, D.~Koutra, and C.~Faloutsos, ``Net-ray: Visualizing and
  mining billion-scale graphs,'' in \emph{Adv in Knowledge Discovery and Data
  Mining}.\hskip 1em plus 0.5em minus 0.4em\relax Springer, 2014, pp. 348--361.

\bibitem{Chakrabarti04netmine:new}
D.~Chakrabarti, Y.~Zhan, D.~Blandford, C.~Faloutsos, and G.~Blelloch,
  ``Netmine: New mining tools for large graphs,'' in \emph{SIAM-DM Workshop on
  Link Analysis}, 2004.

\bibitem{prakash2010eigenspokes}
B.~A. Prakash, A.~Sridharan, M.~Seshadri, S.~Machiraju, and C.~Faloutsos,
  ``Eigenspokes: Surprising patterns and scalable community chipping in large
  graphs,'' in \emph{Adv in Knowledge Discovery and Data Mining}.\hskip 1em
  plus 0.5em minus 0.4em\relax Springer, 2010, pp. 435--448.

\bibitem{Lasalle2511454}
D.~Lasalle and G.~Karypis, ``Multi-threaded graph partitioning,'' in \emph{IEEE
  Int Symp on Parallel and Distributed Processing}, 2013, pp. 225--236.

\bibitem{doi:10.1137/1.9781611973440.11}
D.~Koutra, U.~Kang, J.~Vreeken, and C.~Faloutsos, ``Vog: Summarizing and
  understanding large graphs,'' in \emph{Proc. SIAM Int Conf on Data Mining
  (SDM), Philadelphia, PA}, 2014.

\bibitem{DBLP:conf/icdm/KangF11}
U.~Kang and C.~Faloutsos, ``Beyond 'caveman communities': Hubs and spokes for
  graph compression and mining,'' in \emph{ICDM}, 2011, pp. 300--309.

\bibitem{LeskovecWWW2008}
J.~Leskovec, K.~J. Lang, A.~Dasgupta, and M.~W. Mahoney, ``Statistical
  properties of community structure in large social and information networks,''
  in \emph{WWW}, 2008, pp. 695--704.

\end{thebibliography}

}

\end{document}